\documentstyle[aps,prl,epsf,floats]{revtex}
\tighten
\draft
\begin{document}
 \twocolumn[%
 \hsize\textwidth\columnwidth\hsize\csname@twocolumnfalse\endcsname

\title{Spinons in a Crossed-Chains Model of a 2D Spin Liquid}
\author{Oleg A. Starykh$^1$, Rajiv R. P. Singh$^2$, and Gregory C. Levine$^1$}
\address{$^1$Department of Physics and Astronomy, Hofstra University, Hempstead, NY 11549\\
$^2$Department of Physics, University of California, Davis, CA 95616}
\date{February 11, 2002; cond-mat/0106260} 
\maketitle
\begin{abstract}
Using Random Phase Approximation, we show that a crossed-chains model of a
spin-1/2 Heisenberg spins, with frustrated interchain couplings, has a
non-dimerized spin-liquid ground state in 2D, with deconfined spinons as
the elementary excitations. The results are confirmed by a bosonization study,
which shows that the system is an example of a `sliding Luttinger liquid'.
In an external field, the system develops an incommensurate field-induced
long range order with a finite transition temperature.
\end{abstract}
 ]

Geometrically frustrated magnets have attracted strong interest 
in recent years, primarily because they
provide us with a potentially direct route to the actively sought 
``spin liquid'' 
phases of two- (2d)
and three-dimensional (3d) strongly correlated electron systems.
Begun in the 50's \cite{wannier}, this line of research was dramatically revitalized
by Anderson's proposal \cite{anderson} that such a liquid of spin singlets
is, upon hole doping, adiabatically connected to the
ground state of layered cuprate superconductors.
Whether or not this is so remains to be seen, but 
the following ``question of principle'' ---
can the ground state of a 2d or 3d magnetic system be liquid like? ---
has become one of the most debated questions in condensed
matter physics. Significant insight into this problem has been gained
recently via the dual description of frustrated antiferromagnets
in terms of gauge Ising models \cite{senthil,sachdev,fradkin,readsachdev}.
In this paper we, however, take a more direct, experimentally-motivated
approach, following recent work by Bocquet {\em et al.} \cite{essler} and an earlier
paper by two of us and P. J. Freitas \cite{paradigm}.

The ``spin liquid'' (also known as ``resonating valence bond'', or RVB, phase)
 is defined as 
a liquid of singlet spin pairs covering the lattice, and
is characterized by  the absence of long-range order (LRO),
unbroken spin rotational and  translational symmetries, and
elementary excitations with fractional spin $1/2$ ({\it{spinons}}).
Its physical relevance has been highlighted by the experimental observation of 
spin liquid-like behavior in the
kagom\'e lattice compound SrCr$_{8-x}$Ga$_{4+x}$O$_{19}$ \cite{uemura} and several
pyrochlore antiferromagnets \cite{harris}, e.g. CsNiCrF$_6$. 
Both  materials are based on frustrated units --- triangles
(kagom\'e) and tetrahedra (pyrochlore) --- combined in site sharing arrangements,
which leave the spins in the classical ground-states highly underconstrained \cite{moessner1}.

On the other hand, it is well known that a spin liquid state is
realized in the antiferromagnetic Heisenberg chain (HAFC).
It is then natural to try to build up 2d (or 3d) spin liquid state
from HAFCs - the theoretical investigation of this possibility  is the subject of this paper.

Elementary excitations of the single chain are deconfined gapless spin-1/2 spinons,
which can be visualized as domain walls separating domains of different
orientation (up and down) of staggered
magnetization $\vec{n}(x)=(-1)^x \vec{S}(x)$ along the chain. Consider now
a parallel array of HAFCs coupled by unfrustrated exchange $J_{\perp}$ in the
direction transverse to the chain: no matter how small $J_{\perp}$ is, it immediately leads
to the confinement of spinons because the energy of two domain walls grows
linearly with separation between them, Fig.1a. This is an intuitive reason, confirmed
by detailed calculations \cite{schulz}, for the stabilization of LRO
and appearance of spin-1 magnons (which are bound states of two spinons)
in this situation.
It is clear, however, that this argument fails 
if each spin is coupled, in a transverse direction, to the
zero-spin combination of spins on neighboring chains, {\it{i.e.}} if the transverse coupling
is {\em frustrating}.
Motivated by this simple argument and recent experiments on Cs$_2$CuCl$_4$ \cite{coldea} 
and, especially, Na$_2$Ti$_2$Sb$_2$O \cite{axtell},
we investigate the 
{\em crossed-chains model} \cite{paradigm} (CCM) --- the two-dimensional Heisenberg model
on the lattice shown in Fig.1b,
\begin{equation}
H_{cc}=J\sum_{\langle i,j\rangle}\vec{S}_i \vec{S}_j + J^\prime
\sum_{\langle i,j\rangle}\vec{S}_i \vec{S}_j.
\end{equation}
\begin{figure}[ht]
\epsfxsize=3.in
\centerline{\epsffile{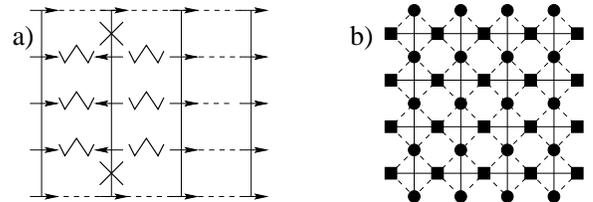}}
\caption{(a) Unfrustrated array of HAFCs. Arrows represent staggered magnetization $\vec{n}(x)$.
Solid (dashed) lines
stand for exchange $J$ ($J_{\perp}$), whereas strings mark broken $J_{\perp}$ bonds.
 Domain walls are indicated by crosses.
(b) The crossed-chains lattice. Filled circles (squares) indicate spins on the
vertical (horizontal) chains. Solid (dashed) lines denote intra-chain $J$ (inter-chain
$J^\prime$) exchange interactions.}
\end{figure}
The intra-chain exchange $J$ couples
neighboring spins in a given row (column) which form a square mesh of HAFC along
the X and Y axes, whereas the inter-chain $J^\prime$ is the nearest
neighbor coupling between spins in rows and columns. The crossed-chain
Hamiltonian interpolates between decoupled HAFCs ($J^\prime=0$),
the 2d pyrochlore lattice ($J^\prime=J$), and $45^\circ$-rotated square lattice
($J=0$)\cite{remark}.
Notice that, when neighboring spins are antiparallel along the chains,
the inter-chain contribution to the effective field
acting on a given spin is zero.

1. {\em RPA analysis of the crossed-chains model  in the weak-coupling limit $J^\prime \ll J$.}
The defining feature of the HAFC is the absence of coherent spin-1 magnon excitations,
already mentioned above. The spectrum of magnetic excitations consists of spin-1/2
noninteracting spinons, and the usual spin-1 magnon is an incoherent two-spinon excitation.
It is then natural to use an approach which properly accounts for this important one-dimensional feature
in the ``decoupled'' limits  $J^\prime/J \rightarrow 0$ and/or $T \gg J^\prime$.
This is provided by a Random-Phase-Approximation (RPA) \cite{scalapino,canada,paradigm,essler}
which has the meaning of
an expansion in the inverse coordination number of the lattice $z_{\perp}$
($=4$ for CCM). Applied to the crossed-chains lattice of Fig.1b,
it gives the following result for the dynamical spin susceptibility of CCM
\begin{eqnarray}
&&\chi_{RPA}(\omega,k_x,k_y)=\Big(\chi_1(\omega, k_x) + \chi_1(\omega, k_y) - 
2 J^{\prime}(\vec{k})\times\nonumber\\
&&\chi_1(\omega, k_x)\chi_1(\omega, k_y)\Big)
\Big(
1 - (J^\prime(\vec{k}))^2\chi_1(\omega, k_x)\chi_1(\omega, k_y)\Big)^{-1}
\label{rpa}
\end{eqnarray}
where momentum $\vec{k}=(k_x, k_y)$ is measured from the antiferromagnetic momentum $(\pi, \pi)$
(lattice spacing is set to unity).
The dynamical susceptibility of a single horizontal ($k_x$) and
vertical ($k_y$) chain, $\chi_1(\omega, k)$,
is known exactly  (for details see, {\em e.g.}, \cite{essler}), 
\begin{eqnarray}
&&\chi_1(\omega, k)=-\frac{\sqrt{\ln(\Lambda/T)}}{2(2\pi)^{3/2} T}
\rho_{\eta}(\frac{\omega-vk}{4\pi T})\rho_{\eta}(\frac{\omega+vk}{4\pi T}),\nonumber\\
&&\rho_{\eta}(x)=\frac{\Gamma(\frac{\eta}{4}-ix)}{\Gamma(1-\frac{\eta}{4}-ix)}
\label{1d}
\end{eqnarray}
where $\Gamma(x)$ is the Gamma function, $\Lambda=24.27 J$ \cite{barzykin} is the high-energy cutoff, 
numerical value of the pre-factor was calculated in \cite{lukyanov},
$\eta=1$ for the isotropic HAFC, 
and $v=\pi J/2$ is the spinon velocity.

The specific lattice structure of CCM is encoded in the Fourier transform
of the inter-chain exchange interaction 
\begin{equation}
J^\prime(\vec{k})=2 J^\prime \sin(k_x/2)\sin(k_y/2),
\label{perp}
\end{equation}
and its frustrating character is clear from the fact that $J^\prime(0,0)=0$.

\begin{figure}[ht]
\epsfxsize=3.in
\centerline{\epsffile{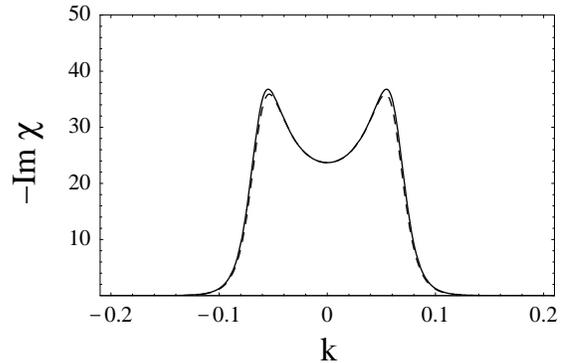}}
\caption{$-\text{Im}\chi_{RPA}(\omega,k,k)$ (solid line) and
$-2\times\text{Im}\chi_1(\omega,k)$ (dashed line) vs. $k$ for 
$J'=J,~\omega=0.1J,~T=0.01J$.}
\label{spec-density}
\end{figure}

\begin{figure}[ht]
\epsfxsize=3.in
\centerline{\epsffile{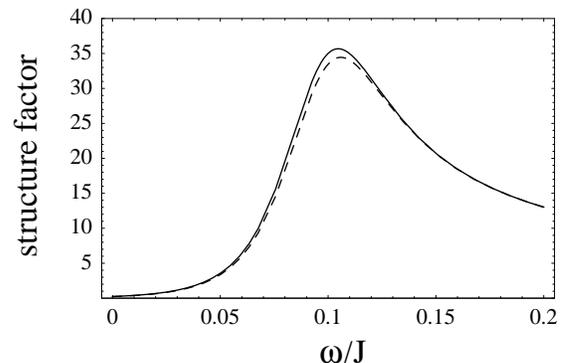}}
\caption{$S_{RPA}(\omega,k)$ (solid line) and $2\times S_1(\omega,k)$ (dashed line) vs. $\omega/J$.
$k=0.06$ and other parameters as in Fig.\ref{spec-density}.}
\label{structure}
\end{figure}

An ordering instability, if any, should show up as a divergence in $\chi_{RPA}(0,\vec{k}_0)$
 at some critical temperature $T_0$ and momentum $\vec{k}_0 =(k_0, \pm k_0)$ along two diagonal
directions where antiferromagnetic fluctuations are the strongest.
Extremizing the denominator of (\ref{rpa}) with respect to $k_0$ one arrives
at the following implicit equation for the ratio $x_0=k_0/(4\pi T_0)$ 
($\Psi(x)$ is the digamma function)
\begin{equation}
\frac{1}{x_0} + \pi \tanh(2\pi x_0) - 2 \text{Im}\Psi(\frac{1}{4} + ix_0)=0.
\label{instability}
\end{equation}
This equation is very similar to the one obtained for the spin model describing 
Cs$_2$CuCl$_4$ \cite{essler} - in that case the coefficient of the first term in (\ref{instability})
is 2 times smaller. This minor difference is, however, extremely important: unlike
the Cs$_2$CuCl$_4$ model, Eq.(\ref{instability}) has {\em no} solution. Thus, RPA analysis
predicts no ordering instability down to, and including, $T=0$: the crossed chains remain
decoupled and provide us with an example of two-dimensional spin liquid
with deconfined spinons as elementary excitations. The ineffectiveness of $J'$ in destabilizing
1d behavior of the crossed-chains is clearly seen from Fig.\ref{spec-density} where we compare 
$\chi_{RPA}(\omega,k,k)$ (\ref{rpa}) with twice the susceptibility of a
single chain $\chi_1(\omega,k)$ (\ref{1d}). Even for unrealistically large $J'=J$ the
difference between these two susceptibilities is hardly observable.
The same is true for corresponding structure factors 
$S_\nu(\omega,k)=-(1-e^{-\omega/T})^{-1}\text{Im}\chi_\nu(\omega,k,k)$ ($\nu=\text{RPA},1$)
which are compared in Fig.\ref{structure}.

The reason for this is purely geometrical and is best illustrated
at $T=0$: the $1/k$ divergence of $\chi_1(\omega=0, k)$, Eq.(\ref{1d}),
is suppressed by the $k^2$ factor from $J^\prime(\vec{k})$, Eq.(\ref{perp}). 
This, again, should be contrasted with the
Cs$_2$CuCl$_4$ lattice where the inter-chain exchange scales as $k$ and fails to suppress
a remaining weakly diverging factor $\sqrt{\ln(1/|k|)}$ due to a marginally irrelevant Umklapp
interaction, which eventually {\em does} lead to
the instability at some incommensurate wavevector \cite{essler}.

Going back to the crossed-chains model it should be noted that the
ordering instability may, in principle, happen at some ``noncritical'' $\vec{k}_0$
in the bulk of the Brillouin zone (BZ) when the ratio $J^\prime/J$ exceeds some critical
value. Analysis of this interesting possibility requires knowledge of $\chi_1(0, k)$ for all
values of momentum inside the BZ, which is currently not available. However,
there exist a closely related model for which the static susceptibility is known exactly
for all momenta -
the Haldane-Shastry chain (HSC) \cite{hsc}. 
At $T=0$ its susceptibility diverges as $1/k$ near the antiferromagnetic momentum 
$\pi$, similar to the HAFC, and remains finite everywhere else,
$\chi_{HS}(0,k)=\arccos|k|/(2\pi v |k| \sqrt{1-k^2})$.
Replacing $\chi_1(0,k)$ in Eq.(\ref{rpa}) with $\chi_{HS}(0,k)$ in order to get an
idea of what may happen in the CCM as the ratio $J^\prime/J$ increases, we find
an instability at $k_0\approx 0.647\pi$ when the ratio of exchanges exceeds
$(J^\prime/J)_{\text{crit}}=(\pi/2)^3 \approx 3.87$. 
This high value of $J^\prime_{\text{crit}}$ in
the HS model on the crossed-chains lattice gives additional support to the robustness
of the spin liquid state.


2. {\em Effect of magnetic field}. Interesting behavior is expected to develop
when the crossed-chains model is subjected to a uniform magnetic field $H$. An applied field
breaks spin-rotational symmetry between transverse and longitudinal susceptibilities \cite{essler}.
The transverse one remains dominated by excitations from the gapless states
at $k=0$ whereas the maximum of the longitudinal susceptibility $\chi_1^{zz}$ shifts to the
incommensurate points $k_{\pm}=\pm \delta(H),~\delta(H)=2\pi M(H)$, where $M(H)$ is the magnetization
of a single HAFC. Since $J^\prime(\vec{k})$ is not affected by the field, the divergence
of $\chi_1^{zz}(0, k\rightarrow k_{\pm})$ is not compensated by inter-chain exchange anymore
and an ordering instability does develop. It signals a transition into an {\em incommensurate}
field-induced ordered phase (FILRO). Both the ordering temperature $T_0$ and
the ordering momentum $\vec{k}_0$ can be calculated from a properly generalized Eq.(\ref{rpa}).
Repeating steps that led to (\ref{instability}) we obtain ($k_0$ is measured from $k_{\pm}$ now)
\begin{eqnarray}
&&\frac{2\pi T_0}{v}\cot(\frac{k_0 \pm \delta}{2}) + \frac{\pi \sinh(2\pi x_0)}{\cosh(2\pi x_0) -
\cos(\frac{\pi}{2\eta})} \nonumber\\
&&- 2 \text{Im}\Psi(\frac{1}{4} + ix_0)=0.
\label{mag-ins}
\end{eqnarray}
This equation has to be solved simultaneously with an equation for $T_0$, 
which is simply the condition that denominator of 
$\chi^{zz}_{RPA}(0,\vec{k}_0)$ is equal to zero. Notice that 
both the scaling exponent $\eta(H)$ and the spinon velocity $v(H)$ decrease with magnetic field,
 in particular $\eta(H)=1/2$ for saturating field $H \geq 2 J$ \cite{essler}.
However, analytical solution is still possible in the limit of weak applied field when
$\delta(H) \rightarrow 0$ thanks to the following inequalities $v k_0 \ll T_0, ~k_0 \ll \delta$
which can be checked {\em a posteriori}.
We find {\em incommensurate} ordering with $k_{0,\pm}= \pm c_1\delta^{(2\eta+1)/(2\eta -1)}$
developing at 
$T_0=c_2 v (\delta \sqrt{J^\prime/v})^{2\eta/(2\eta -1)}$, where $c_{1,2}$ are weakly $H$-dependent
constants of order 1. So that for $H\rightarrow 0$ the scaling is 
$k_{0,\pm}\sim \pm \delta^3, ~T_0 \sim J^\prime \delta^2$.
Notice that 2d ordering occurs at momenta different from 
$k_{\pm}$ at which susceptibility has a maximum for independent chains,
although the difference is probably too small to be observed experimentally.
Observation of such a FILRO in materials such as Na$_2$Ti$_2$Sb$_2$O would be a
direct signature for the applicability of the coupled-chains model presented here.

3. {\em Scaling analysis: sliding Luttinger liquid phase.}

We now employ  bosonization to investigate effect of the inter-chain interaction $V$
 at finite temperature $T$. We consider the mesh made of equal number of horizontal ($h$)
and vertical ($v$) chains, $N_h=N_v=N$, each of length $L=Na$, where $a$ is the
lattice spacing. 
In the continuum limit each site spin is represented by a sum of uniform $\vec{J}$ and staggered
$\vec{n}$ fields \cite{gnt}. In addition, $J_{v/h}^\nu=J_{v/h,R}^\nu + J_{v/h,L}^\nu$,
where $J_{R/L}^\nu$ are chiral WZW currents.
The inter-chain interaction reads
\begin{equation}
V=g\int dx dy \vec{J}_h(x,y)\cdot \vec{J}_v(x,y)
\label{VV}
\end{equation}
where $g=2J'$. 
Note that due to the geometry of the problem the staggered magnetization
$\vec{n}$, which has scaling dimension $\Delta(\vec{n})=1/2$, 
does not show up in this equation. As a result, the
scaling dimension of the integrand in (\ref{VV}) is $2$.
The bare Hamiltonian $H_0$ describes total of $2N$ independent vertical and horizontal chains.
From here follows the defining feature of our model: $J-J$ correlations are non-zero
only for like currents (i.e. {\em h-h} and {\em v-v}) of the same chirality 
(i.e. R-R and L-L) which belong to the {\em same chain}, e.g.
\begin{eqnarray}
&&\langle J^\mu_{v,R}(x,y,\tau) J^\nu_{v,R}(0,0,0)\rangle = a \delta(x) \delta_{\mu,\nu}
(\pi T/v)^2 
\nonumber\\
&&\times \frac{1}{8\pi^2} \{\frac{1}{\sin^2[\pi T(v\tau + i y)/v]} +
\frac{1}{\sin^2[\pi T(v\tau - i y)/v]}\}
\label{G}
\end{eqnarray}
and $\langle J_h J_h \rangle$ is obtained by replacing $v \rightarrow h,~y \leftrightarrow x$.
 The `same-chain' condition is contained in an important
delta-function  which appears with pre-factor $a$ because
$\delta(x-x')=\delta_{n,n'}/a$. Conformal invariance of $H_0$ was used to write (\ref{G})
at finite $T$ \cite{nmr,gnt}.

Our idea is to consider corrections to the bare free energy of $2N$ chains
$F_0/L^2 = \pi v/(2 a^3) - \pi T^2/(3 v a)$
(momentum cut-off $\Lambda=\pi/a$ was used to get the first term).
First correction is $\delta F^{(2)}=-T\langle V^2 \rangle /2$ where
\begin{eqnarray}
\langle V^{2} \rangle =&&(ga)^2 \int dx dy \int_0^{1/T} d\tau d\tau' 
\langle J_h(x,y,\tau) J_h(x,y,\tau')\rangle 
\nonumber\\
&&\times \langle J_v(x,y,\tau)J_v(x,y,\tau')\rangle
\label{g^2}
\end{eqnarray}
Notice that due to delta-functions in (\ref{G}) spacial coordinates of currents in 
(\ref{g^2}) are forced to {\em coincide}. Short-distance divergence of the integrand
in (\ref{g^2}) is regularized by $a$, leading to
\begin{equation}
\langle V^{2} \rangle= \Big(\frac{gaLT}{4v^2}\Big)^2
\int_0^{2\pi} \frac{d\phi}{2\pi [\cos\phi - \cosh(2\pi T a/v)]^2}
\end{equation}
The resultant correction to the free energy is {\em irrelevant},
\begin{equation}
\delta F^{(2)}/L^2 =  -\frac{\pi v}{2a^3} \Big(\frac{ga}{8\pi^2 v}\Big)^2 
\{1 - \frac{1}{15}(2\pi T a/v)^4 \}.
\end{equation}
This  can be understood as follows. In the weak-coupling limit, where
each crossing can be treated independently from others, interaction $V$ acts at a {\em point}
rather than along the entire length of two chains forming that particular crossing.
This is the physical reason for the absence of integration over the {\em relative}
spatial coordinates in (\ref{g^2}). It is well known that for the point interaction
marginal dimension separating relevant from irrelevant perturbations is $1$ \cite{kane}.
Since $\Delta(V)=2 >1$, $V$ is {\em irrelevant}.

At this order the ground state of the CCM is that of {\em decoupled} spin chains.
At $T=0$ there are no correlations between spins on different
chains due to irrelevancy of $V$ but at finite $T$ weak inter-chain correlations with
a square $C_{4v}$ symmetry will be present \cite{lubensky}, in qualitative agreement with our
RPA expression (\ref{rpa}).
Dimensional estimate of higher order in $g$ corrections  show that their contribution
is {\em at most} marginal. This allows us to identify the CCM as a $SU(2)$ invariant
{\em spin liquid} with deconfined spinons as elementary excitations.
Following earlier works on models with $U(1)$ symmetry 
\cite{sliding,lubensky}
it can be called a {\em sliding Luttinger liquid}.  

We thank A. G. Abanov, L. Balents, F. Essler, A. Furusaki, M. P. A. Fisher, C. Kane,
S. Sachdev and A. M. Tsvelik for helpful discussions and suggestions.
This research is supported by ITP (Santa Barbara) Scholarship and 
an award from Research Corporation (O.A.S.)
and by NSF grant DMR9986948 (R.R.P.S.).

{\bf Note added:} Since the submission of our work numerical studies have further addressed
the $J'=J$ limit of the model (the 2d pyrochlore lattice) \cite{fouet}. They find a 
valence-bond crystal phase with a spin-gap. These results are not in contradiction with
our findings. In fact, making the reasonable assumption that the ground state energy
in the valence-bond crystal phase and the sliding Luttinger liquid phase is weakly dependent
on $J'/J$, leads to an estimate for a quantum phase transition between the two at
$J'/J =0.85$. This allows for a wide range $0< J'/J < 0.85$ of values where the sliding
Luttinger liquid phase could exist.

\end{document}